\documentstyle[11pt,aaspp4]{article}

\input psfig

\begin{document}

\title{The Importance of Einstein Rings\footnote{Based on
Observations made with the NASA/ESA Hubble Space Telescope, obtained at
the Space Telescope Science Institute, which is operated by AURA, Inc.,
under NASA contract NAS 5-26555.}}

\author{C.S. Kochanek$^{(a)}$, C.R. Keeton$^{(b)}$ and B.A. McLeod$^{(a)}$}

\affil{$^{(a)}$Harvard-Smithsonian Center for Astrophysics, 60 Garden St.,
  Cambridge, MA 02138}
\affil{email: ckochanek, bmcleod@cfa.harvard.edu}
\affil{$^{(b)}$Steward Observatory, University of Arizona, 933 N. Cherry Ave.,
   Tucson, AZ 85721}
\affil{email: ckeeton@as.arizona.edu}

\begin{abstract}
We develop a theory of Einstein rings and demonstrate it using the infrared
Einstein ring images of the quasar host galaxies observed in PG~1115+080,
B~1608+656 and B~1938+666.  The shape of an Einstein ring accurately
and independently determines the shape of the lens potential and the shape of
the lensed host galaxy.  We find that the host galaxies of PG~1115+080, 
B~1608+656 and B~1938+666 have axis ratios of $0.58\pm0.02$, $0.69\pm0.02$ 
and $0.62\pm0.15$ including the uncertainties in the lens models.  The Einstein 
rings break the degeneracies in the mass distributions or Hubble constants 
inferred from observations of gravitational lenses.  In particular, the 
Einstein ring in PG~1115+080 rules out the centrally concentrated mass 
distributions that lead to a high Hubble constant 
($H_0>60$~km~s$^{-1}$~Mpc$^{-1}$) given the measured time delays.  
Deep, detailed observations of Einstein rings will be revolutionary for 
constraining mass models
and determining the Hubble constant from time delay measurements.   
\end{abstract}

\keywords{cosmology: gravitational lensing; distance scale: Hubble constant}

\section{Introduction}

Gravitational lenses are excellent tools for studying the gravitational potentials
of distant galaxies and their environments.  Even the simplest models can measure
some properties of the lens, such as the mass inside the Einstein ring, with an
accuracy far beyond that possible with any other astrophysical 
probe of galaxy mass distributions (see, e.g., Kochanek 1991).  More detailed explorations of the potentials,
particularly determinations of the radial mass distribution and the interpretation of
time delays for estimating the Hubble constant, are limited by 
the number of constraints supplied by the data (see, e.g., Impey et al. 1998, Koopmans \& Fassnacht
1999, Williams \& Saha 2000, Keeton et al. 2000 for recent examples).  
In fact, all well-explored lenses with time delays currently
have unpleasantly large degeneracies between the Hubble constant and the structure
of the lensing potential.  Most of the problem is created by the aliasing between
the moments of the mass distribution we need to measure (primarily the monopole)
and higher order moments (quadrupole, octopole, $\cdots$) when we have constraints
at only two or four positions.  Progress requires additional model constraints
spread over a wider range of angles and distances relative to the lens center.

Many gravitational lenses produced by galaxies now contain multiple 
images of extended sources.  Radio lenses, even if selected to be dominated by compact cores, 
commonly show rings and arcs formed from multiple images of the radio lobes and jets associated
with the AGN (see, e.g., Hewitt et al. 1988, Langston et al. 1989, Jauncey et al. 1991,
Lehar et al. 1993, Lehar et al. 1997, King et al. 1997).  More importantly, short infrared
observations of gravitational lenses with the Hubble Space Telescope frequently find Einstein
ring images of the host galaxies of the multiply-imaged quasars and radio sources (e.g.
King et al. 1998, Impey et al. 1998, Kochanek et al. 2000, Keeton et al. 2000).  Since all
quasars and radio sources presumably have host galaxies, we can {\it always} find an 
Einstein ring image of the host galaxy given a long enough observation.  
Moreover, finding a lensed host galaxy is easier than finding an unlensed host
galaxy because the lens magnification enormously improves the
contrast between the host and the central quasar.

The Einstein ring images of the host galaxy should provide the extra constraints needed 
to eliminate model degeneracies.  In particular, complete Einstein rings should almost
eliminate the aliasing problem.  The first step towards using the host galaxies as 
constraints is to better understand how Einstein rings constrain lens models. In this paper 
we develop a general theory for the shapes of Einstein rings and demonstrate its utility
by using it to model the Einstein ring images of the quasar host galaxies in PG~1115+080 
(Impey et al. 1998), B~1608+656 (Fassnacht et al. 1996) and 
B~1938+666 (King et al. 1998).  In \S2 we briefly review the theory of gravitational lensing and 
then develop a theory of Einstein rings in \S3.  We apply the models to the three
lenses in \S4 and summarize our results in \S5.

\def\vx{\vec{x}}
\def\vu{\vec{u}}
\def\mt{{\bf M}}

\section{Standard Lens Theory and Its Application to Lens Modeling}

In \S2.1 we present a summary of the basic theory of lenses as reviewed
in more detail by Schneider, Ehlers \& Falco (1992).  In \S2.2 we review the
basic procedures for fitting lenses composed of multiply-imaged point sources
and define the lens models we will use.  

\subsection{A Quick Review}

We have a source located at $\vu$ in the source plane, which 
is lensed in a thin foreground screen by the two-dimensional potential $\phi(\vx)$.  
The images of the source are located at solutions of the lens equation
\begin{equation}
   \vu = \vx - \nabla_{\vx} \phi(\vx)
\end{equation}
which will have 1, 2 or 4 solutions for standard lens models.  In the multiple image
solutions, one additional image is invisible and trapped in the singular core of the galaxy.
If the source is extended, the images are tangentially stretched into arcs around
the center of the lens, merging into an Einstein ring if the source is large
enough.  The local distortion of an image is determined by the inverse magnification 
tensor
\begin{equation}
    \mt^{-1} = \left(
             \begin{array}{cc}
                 1-\phi,_{xx}     &-\phi,_{xy} \\
                 -\phi,_{xy}      &1-\phi,_{yy}
             \end{array}
             \right)
\end{equation}
whose determinant, $M^{-1}$, sets the relative flux ratios of unresolved images.  
Surface brightness is conserved by lensing, so a source with surface brightness
$f_S(\vu)$ produces an image with surface brightness $f_I(\vx)=f_S(\vu)$ where the 
source and lens coordinates are related by eqn. (1).  Real data has finite
resolution due to the PSF $B$, so the observed image is the convolution
$f_D(\vx) = B * f_I(\vx)$ rather than the true surface brightness $f_I(\vx)$.
The observed image can be further modified by extinction in the optical/infrared
and Faraday rotation for polarization measurements in the radio.

\subsection{Standard Modeling Methods and Problems} 

The procedures for modeling lenses consisting of unresolved point sources are
simple to describe and relatively easy to implement.  Given a set of observed
image positions $\vx_i$, and a lens model predicting image positions $\vx_i(\vu)$,
the goodness of fit is measured with a $\chi^2$ statistic
\begin{equation}
      \chi^2_{point} = \sum_i { |\vx_i -\vx_i(\vu)|^2 \over \sigma_i^2 } 
\end{equation}  
with (in this case) isotropic positional uncertainties $\sigma_i$ for each image.
The statistic is minimized with respect to the unobserved source position, $\vu$, and the 
parameters of the lens model, where the model parameters may be further constrained 
by the observed properties of the lens (position, shape, orientation $\cdots$). 
The relative image
magnifications are also constrained by the relative fluxes, but the uncertainties
are dominated by systematic errors in the fluxes due to extinction,
temporal variations, 
microlensing and substructure rather than measurement uncertainties.  If the images have signed
fluxes $f_i$ and the source has flux $f_S$, then we model the fluxes with
another simple $\chi^2$ statistic
\begin{equation}
      \chi^2_{flux} = \sum_i { (f_i - |M_i| f_S)^2 \over \sigma_i^2 } 
\end{equation}  
where $M_i$ is the magnification at the position of the image and $\sigma_i$ is the
uncertainty in the flux $f_i$.

We fit the lenses discussed in \S4 using models based on the softened isothermal 
ellipsoid, whose surface density in units of the critical surface density is
$\kappa_{IE}(m^2,s) = (b/2) (s^2+m^2)^{-1/2}$ for core radius $s$, ellipsoidal
coordinate $m^2=x^2+y^2/q_l^2$, and axis ratio $q_l=1-e_l$.  The model parameters
are the critical radius $b$, the core radius $s$, the ellipticity $e_l$ and the
position angle, $\theta_l$, of the major axis (see Kassiola \& Kovner 1993,
Kormann, Schneider \& Bartelmann 1994, Keeton \& Kochanek 1998). We use either the
singular isothermal ellipsoid (SIE), $\kappa_{IE}(m^2,0)$, in which the core 
radius $s\equiv0$, or
the pseudo-Jaffe model, $\kappa(m^2) =\kappa_{IE}(m^2,0)-\kappa_{IE}(m^2,a)$,
which truncates the mass distribution of the SIE at an outer truncation radius
$a$ (see Keeton \& Kochanek 1998).
This allows us to explore the effects of truncating the mass distribution
on the lens geometry.  We also allow the models to be embedded in an external
tidal shear field characterized by amplitude $\gamma$ and  position angle
$\theta_\gamma$, where the angle is defined to point in the direction from
which an object could generate the shear as a tidal field.

The simplest possible model of a lens has 5 parameters (lens position, mass,
ellipticity, and orientation).  Realistic models must add an independent 
external shear (see Keeton, Kochanek \& Seljak 1997), and additional 
parameters (2 in most simple model families) describing the radial mass distribution 
of the lens galaxy, for a total of 9 or more parameters.  A two-image lens supplies
only 5 constraints on these parameters, so the models are woefully underconstrained.
A four-image lens supplies 11 constraints, but the 3 flux-ratio constraints are
usually weak constraints because of the wide variety of systematic problems in
interpreting the image fluxes (extinction, time variability, microlensing, local perturbations
$\cdots$, see Mao \& Schneider 1998).  Much of the problem is due to aliasing, 
where the quadrupole in particular, but also the higher angular multipoles, can
compensate for large changes in the monopole structure given the limited 
sampling provided by two or four images. The paucity of constraints leads to the large 
uncertainties in Hubble constant estimates from time delay measurements (see Impey et al. 1998,
Koopmans \& Fassnacht 1999, Williams \& Saha 2000, Keeton et al. 2000, Zhao
\& Pronk 2000, Witt, Mao \& Keeton 2000 for a wide variety of examples).  The
solution is to find more constraints. 
 
\begin{figure}
\centerline{\psfig{figure=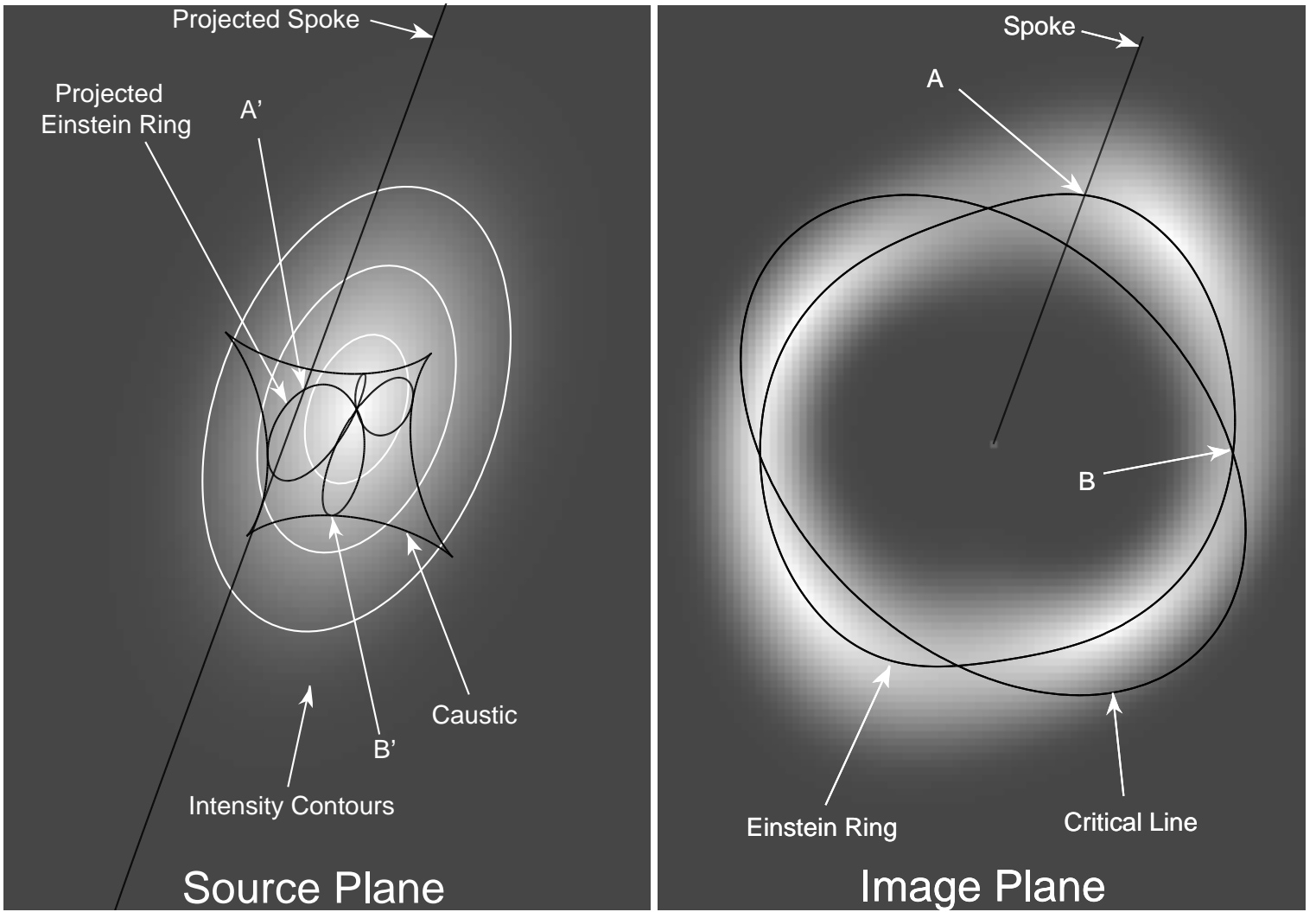,width=8.0in}}
\caption{An illustration of ring formation by an SIE lens.  An ellipsoidal source (left grayscale) is lensed into
an Einstein ring (right grayscale).  The source plane is magnified by a factor of 2.5 relative
to the image plane.   The tangential caustic (astroid on left) and critical line
(right) are superposed.  The Einstein ring curve is found by looking for the peak brightness 
along radial spokes in the image plane.  For example, the spoke in the illustration defines
point A on the ring curve.  The long line segment on the left is the projection of the spoke
onto the source plane.
Point A on the image plane corresponds to point A$'$ on the source plane where the projected spoke 
is tangential to the intensity contours of the source.  The ring in the image plane projects into 
the four-lobed pattern on the source plane.  Intensity maxima along the ring correspond to the
center of the source.  Intensity minima along the ring occur where the ring crosses the
critical line (e.g. point B).  The corresponding points on the source plane (B$'$) are 
where the astroid caustic is tangential to the intensity contours.
  }
\end{figure}

\begin{figure}
\centerline{\psfig{figure=fig2a.ps,height=3.5in}\psfig{figure=fig2b.ps,height=3.5in}}
\centerline{\psfig{figure=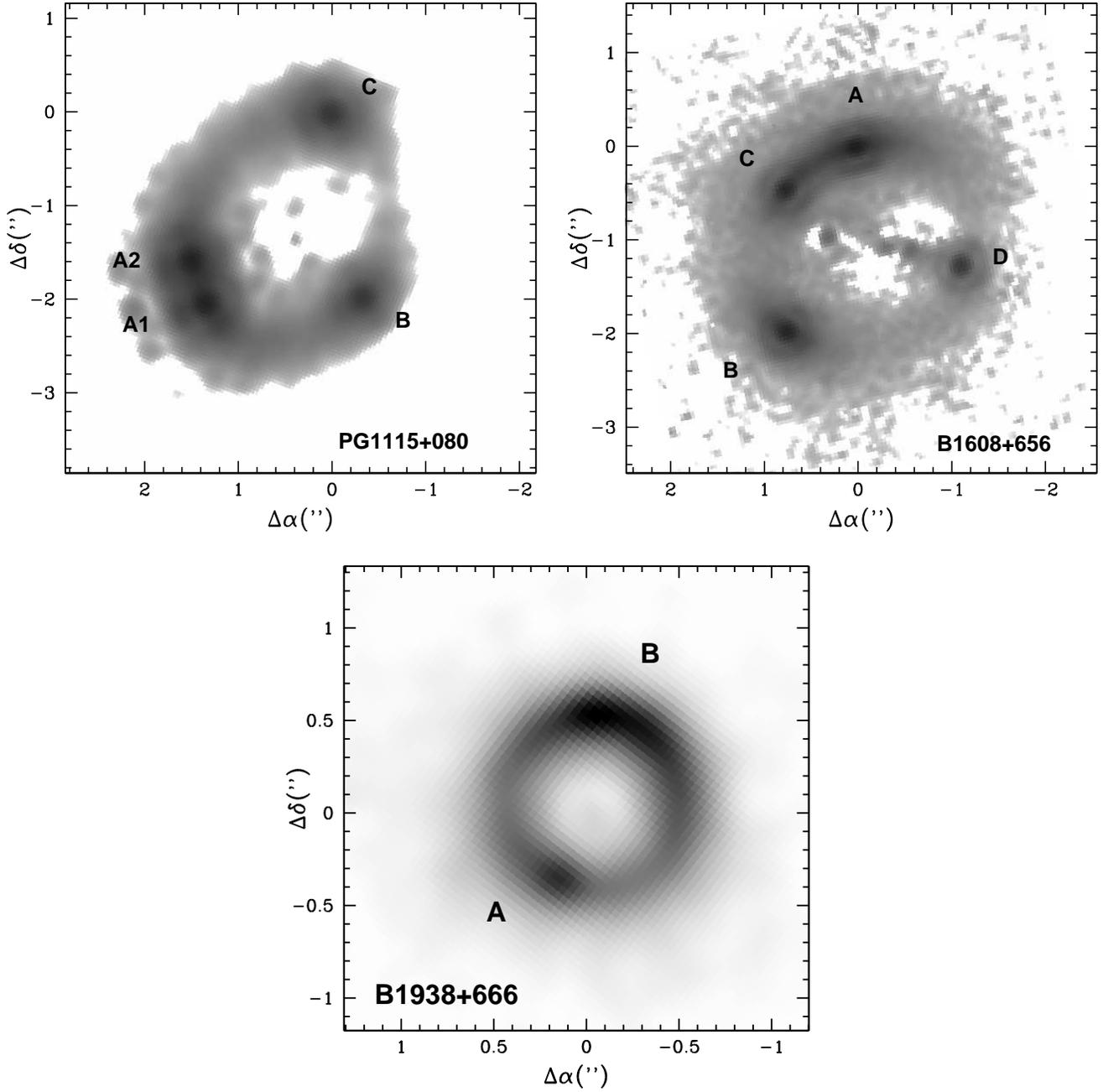,height=3.5in}}
\caption{The CASTLES (Falco et al. 2000) HST/NICMOS H-band images  PG~1115+080 (top left), B~1608+656 (top right),
  and B~1938+666 (bottom).  We partially subtracted the bright point sources in PG~1115+080 and
  B~1608+656 and completely subtracted the lens galaxies for all three systems to better show
  the ring structure. 
  }
\end{figure}

\begin{figure}
\centerline{\psfig{figure=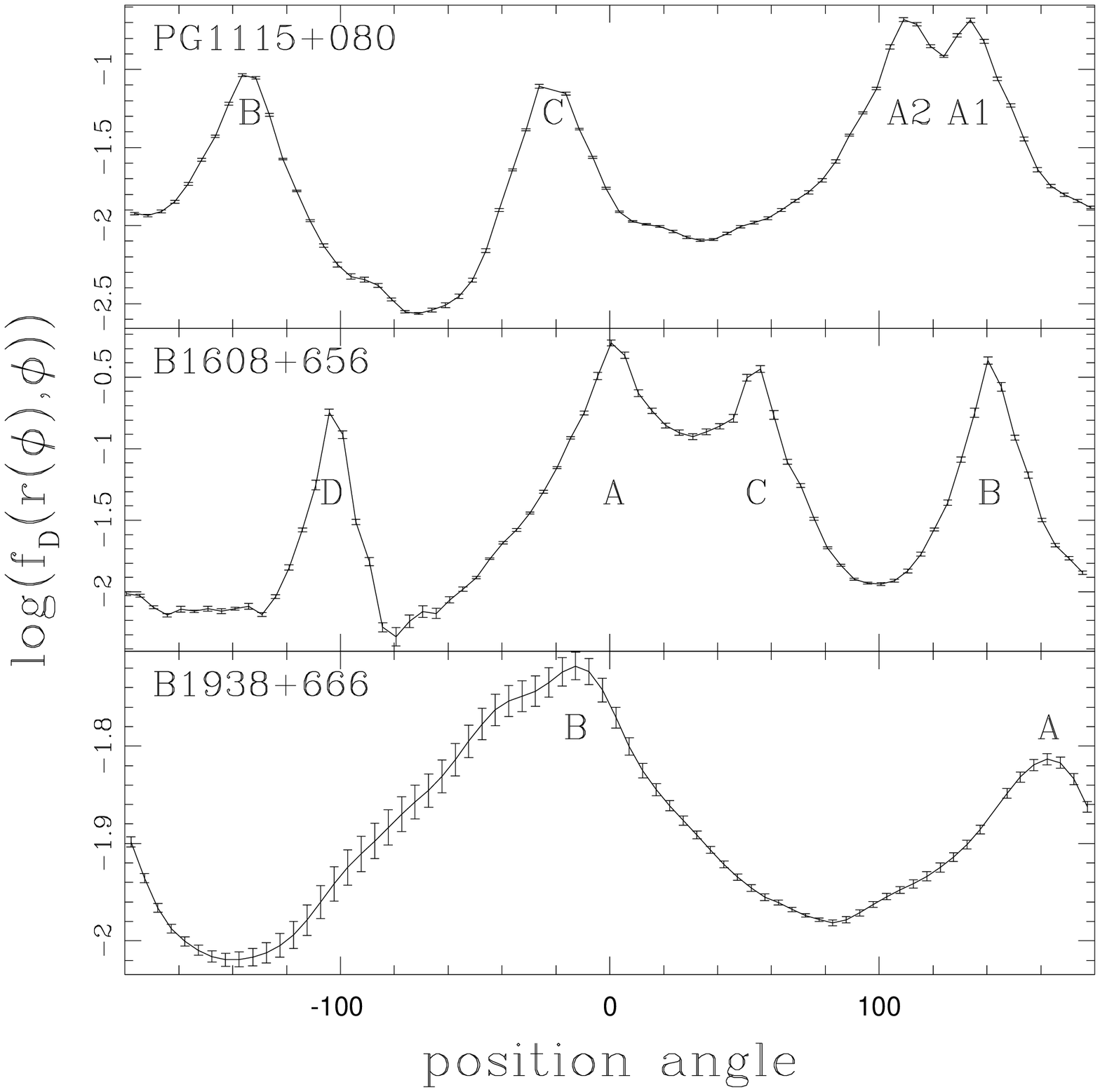,height=5.5in}}
\caption{The ring brightness profiles as a function of the spoke position angle.  We show 
  $\log f_D(r(\phi),\phi)$ for PG~1115+080 (top), B~1608+656 (middle) and
  B~1938+666 (bottom) after supressing the point sources in PG~1115+080 and B~1608+656.
  The (labeled) maxima constrain the host center and the minima constrain where the critical line
  crosses the Einstein ring.  The properties of the extrema were determined using parabolic
  fits to the local extrema of the curves.  The two local minima in the profile between the
  B and D images of B~1608+656 are an artifact due to noise, extinction or lens galaxy
  subtraction errors.
  }
\end{figure}

\section{A Simple Quantitative Theory of Einstein Rings}

Einstein rings are one of the most striking properties of lenses.  Just as 
they are visually striking, we can easily measure some of their properties.
In \S3.1 we discuss the shape of the curve defined by the peak surface 
brightness of an Einstein ring as a function of azimuth.  We need a new theory
because the curve is a {\it pattern} that cannot be modeled using the
standard methods for fitting multiply-imaged point sources.  In \S3.2
we discuss the locations of the maxima and minima of the brightness of
an Einstein ring.  Finally, in \S3.3 we describe the statistic we use 
to fit Einstein rings and its relation to observations.  Figure 1 
illustrates how an Einstein ring is formed, and Figures 2 and 3 show the
real examples we will consider in \S4.

\subsection{The Ring Curve}

We visually trace the ring as the curve formed by the peak surface brightness 
of the ring around the lens galaxy.  We find the curve by locating the peak 
intensity on radial spokes in the image plane, $\vx(\lambda) = \vx_0 + \lambda(\cos\phi,\sin\phi)$,
which are parameterized by $\lambda$ and originate from an arbitrary point 
$\vx_0$ near the center of the ring (see Figure 1). More complicated 
parameterizations are 
possible but unnecessary.  For each azimuth $\phi$ we determine the position
of the maximum, $\lambda(\phi)$, and the image flux at the maximum.
Mathematically, the extrema of the surface brightness of the image $f_D(\vx)$ along 
the spoke are the solutions of
\begin{equation}
   0 = \partial_\lambda f_D(\vx) = \nabla_{\vx} f_D (\vx) \cdot {d \vx \over d \lambda }.
\end{equation} 
The next step is to translate the criterion for the ring location into the source plane.
For simplicity, we assume that the image is a surface brightness map so
that $f_S(\vu)=f_I(\vx)=f_D(\vx)$.  Using the lens equation (1) and the definition
of the magnification tensor (eqn. (2)), the criterion for the maximum becomes 
\begin{equation}
       0 = \nabla_{\vu} f_S (\vu) \cdot \mt^{-1} \cdot {d \vx \over d \lambda }.
\end{equation} 
Geometrically, we are finding the point where the tangent vector of the curve projected
onto the source plane ($M^{-1} \cdot d\vx/d\lambda$) is perpendicular to the local gradient 
of the surface brightness ($\nabla_{\vu} f_S (\vu)$), as illustrated in Figure 1. 

\def\st{{\bf S}}
While the structure of the lensed image is complicated, in many systems it is 
reasonable to assume that the structure of the source has some regularity and 
symmetry.  We require a model for the surface brightness of the source over a 
very limited region about its center where almost all galaxies can be 
approximated by an ellipsoidal surface brightness profile.  We assume that 
the source has
a surface brightness $f_S(m^2)$ which is a monotonically decreasing function,
$d f_S/dm^2 < 0$, of an ellipsoidal coordinate $m^2=\Delta\vu \cdot \st\cdot  \Delta\vu$.  The
source is centered at $\vu_0$, with $\Delta\vu=\vu-\vu_0$, and its shape is described
by the two-dimensional shape tensor $\st$.\footnote{An ellipsoid is described by an axis
ratio $q_s=1-e_s<1$ and major axis position angle $\theta_s$.  In the principal
axis frame the shape tensor $\st_D$ is diagonal with components $1$ and $q_s^{-2}$.  
The shape tensor $\st=R^{-1}\st_D R$ 
is found by rotating the diagonal shape tensor with the rotation matrix $R(\theta_s$)
corresponding to the major axis PA of the source $\theta_s$. }
A non-circular source is essential to explaining the observed rings.  
With these assumptions, the location of the ring depends only the shape of the 
source and not on its radial structure. The position
of the extremum is simply the solution of
\begin{equation}
       0 = \Delta\vu \cdot \st \cdot \mt^{-1} \cdot {d \vx \over d \lambda }.
\end{equation}
The ring curve traces a four (two) lobed cloverleaf
pattern when projected onto the source plane if there are four (two) images 
of the center of the source (see Figure 1).  These lobes touch the tangential 
caustic at their maximum ellipsoidal radius from the source center, and these
cyclic variations in the ellipsoidal radius produce the brightness variations 
seen around the ring (see Figure 3, \S3.2). 

Two problems affect using eqn. (7) to define the ring curve.  First,
it assumes that the image is a surface brightness map.  In practice,
$f_D(\vx) = B * f_I(\vx) \neq f_I(\vx)$, so the ring location determined
from the data can be distorted by the PSF.  One major advantage of fitting
the ring curve is that it is insensitive to the effects of the PSF --
the PSF distorts the ring 
location only when there is a strong flux gradient along the ring.
We see this primarily when the ring contains much brighter point sources, 
and the problem can be mitigated by fitting and subtracting most of the flux 
from the point sources before measuring the ring position.  Alternatively
we can use deconvolved images.   
Monte Carlo experiments 
with a model for the ring are necessary to determine where the ring curve is biased from the
true ring curve by the effects of the PSF, but the usual signature is
a ``ripple'' in the ring near bright point sources (as seen between
the A1 and A2 images in Figure 4).  Second, we are assuming
that the source is ellipsoidal, although the model could be generalized 
for more complicated source structures.

We can illustrate the physics governing the shapes of Einstein rings using
the limit of a singular isothermal ellipsoid (SIE) for the lens.  The 
general analytic solution for the Einstein ring produced by an ellipsoidal 
source lensed by a singular isothermal galaxy in a misaligned external shear 
field is presented in Appendix A.  We use coordinates 
($x=r\cos\theta$, $y=r\sin\theta$) centered on the lens galaxy.  
The SIE (see \S2.2) has a critical radius scale $b$, ellipticity $e_l=1-q_l$, 
and its major axis lies along the $x$-axis. We add an external shear field characterized by
amplitude $\gamma$ and orientation $\theta_\gamma$.  The source is an ellipsoid with 
axis ratio $q_s=1-e_s$ and major axis angle $\theta_s$ located at a position of 
($\rho\cos\theta_0$, $\rho\sin\theta_0$) from the lens center. 
The tangential critical line of the model is 
\begin{equation}
     r_{crit}/b =   1 + { e_l \over 2 }\cos 2\theta - \gamma \cos 2(\theta-\theta_\gamma)
\end{equation}
when expanded to first order in the shear and ellipticity.  If we expand the
solution for the Einstein ring to first order in the shear, lens ellipticity, source
ellipticity and source position ($\rho/b \sim \gamma \sim \epsilon_l$ in a four-image
lens), the Einstein ring is located at
\begin{equation}
    r_E/b = 1 + { \rho \over b } \cos(\theta-\theta_0)
        - { e_l \over 6 }\cos 2\theta + \gamma \cos 2(\theta-\theta_g).
\end{equation}
At this order, the average radius of the Einstein ring is the same as that of the
tangential critical line.  The quadrupole of the ring has a major axis orthogonal
to that of the critical line if the shear and ellipticity are aligned, $e_l=0$ 
or $\gamma=0$, but can be misaligned in the general solution because of the
different coefficients for $e_l$ in eqns. (8) and (9).  For an external shear, the ellipticity of the ring
is the same as that of the critical line, while for the ellipsoid the ring is
much rounder than the critical line.  The ring has a dipole moment when expanded
about the center of the lens, such that the average ring position is the
source position.  At this order, the ring shape appears not to depend on the source
shape.      
  
The higher order multipoles of the ring are important, as can be seen from the very 
non-ellipsoidal shapes of many observed Einstein rings (see Figure 2).  The even multipoles 
of the ring are dominated by the shape of the lens potential while the odd multipoles are 
dominated by the shape of the source (see eqn. A5).  Their large amplitudes are driven by 
ellipticity of the source, because the ellipticity of the lens potential is 
small ($\gamma \sim e_l/3\sim 0.1$) even if the lens is flattened ($e_l\sim0.5$), while 
the ellipticity of the source is not ($e_s \sim 0.5$).
For example, in a circular lens ($e_l=0$, $\gamma=0$) the ring is located at 
\begin{equation}
     { r_E \over b } = 1 + { \rho \over b } \left[ 
  { (2-e_s) \cos (\theta-\theta_0) + e_s \cos(2\phi_0-\theta-\theta_0) \over
           2 -e_s + e_s \cos 2(\phi_0-\theta) } \right]
\end{equation}
which has only odd terms in its multipole expansion and converges slowly for
flattened sources.  The ring shape is a weak function of the source shape only if 
the potential is nearly round and the source is almost centered on the lens.  

\subsection{Maxima and Minima in the Brightness of the Ring}

The other easily measured quantities for an Einstein ring are the locations
of the maxima and minima in the brightness along the ring.  Figure 3 shows
the brightness profiles as a function of the spoke azimuth for the three
lenses.  The brightness profile is $f_I(r(\phi),\phi)$ for a spoke at 
azimuth $\phi$ and radius $r(\phi)$ determined from eqn. (7), and it 
has an extremum when $\partial_\phi f_I=0$.  For the ellipsoidal model
there are maxima at the images of the center of the host galaxy
($\Delta\vu=0$), and minima when the ring crosses a critical line and the 
magnification tensor is singular ($|M^{-1}|=0$).  In fact, these are general 
properties of Einstein rings and do not depend on our assumption of ellipsoidal 
symmetry (see Blandford, Surpi \& Kundic 2000).  The extrema at the critical
line crossings are created by having a merging image pair on the critical
line.  The two merging images are created from the same source region and
have the same surface brightness as the source, so the surface brightness 
of the ring must be continuous across the critical line.   Alternatively,
the lobed pattern traced by the ring curve on the source plane (see Figure 1)  
touches the tangential caustic at maxima of the ellipsoidal radius from 
the source center.  The points where the lobes touch the caustic on the
source plane correspond to the points where the ring crosses the critical
line on the image plane, so the ring brightness profile must have a 
minimum at the crossing point.

We can accurately measure the locations of the maxima, since they correspond to the
center of the host galaxy or the positions of the quasars.  The positions of the minima, 
the two or four points where the Einstein ring crosses the tangential critical line, are 
more difficult to measure accurately.  First, if the minimum lies between a close pair of 
images, like the one between the bright A1/A2 quasar images in PG~1115+080 (see Figures 2 and 3),
the wings of the PSF must be well modeled to measure the position of the minimum
accurately.  Second, the high tangential magnifications near crossing points reduce
the surface brightness gradients along the ring near the minima. The smaller the gradient, 
the harder it is to accurately measure the position of the flux minimum. Third, the
minima will be more easily affected by dust in the lens than the sharply peaked
maxima.  Clean, high precision data is needed to accurately measure the crossing 
points.

\subsection{Modeling Method}

The final statistic for estimating the goodness of fit for the ring model has 
three terms.  The first term is the fit to the overall ring curve. We measure 
the ring position $\lambda_o$ and its uncertainty $\sigma_\lambda$ for a series
of spokes radiating from a point near the ring center with separations at the 
ring comparable to the resolution of the observations so that they are 
statistically independent.  Given a lens model and the 
source structure ($\vu_0$, $\theta_s$ and $q_s=1-e_s$) we solve eqn. (7) to derive the 
expected position of the ring $\lambda_m$, and add 
$(\lambda_o-\lambda_m)^2/\sigma_\lambda^2$ to our fit statistic 
for each spoke.  The second term
compares the measured position of the ring peaks to the positions predicted from
the position of the host galaxy $\vu_0$.  We simply match the predicted and
observed peak positions using a statistic similar to eqn. (3).  If we also
fit the separately measured quasar positions, we must be careful not to double
count the constraints.  At present, we have used a larger uncertainty for
the position of the ring peaks than for the positions of the quasars so that
the host center is constrained to closely match that of the quasar without
overestimating the constraints on the position of the quasars and host 
center.  Finally, we determine where the observed ring crosses the model
critical line and compare it to the positions of the flux minima and their
uncertainties, again using a statistic similar to eqn. (3).  Because we use
the observed rather than the model ring to estimate the positions of the 
critical line crossings, the constraint is independent of any assumptions
about the surface brightness profile of the source.

We can estimate the measurability of the parameters from simple considerations
about the measurement accuracy.  We measure the ring radius at $N$ independent points 
measured around the ring with uncertainties of $\sigma_r$ in the radius of each point.  
If we measure the multipole moments of the ring,\footnote{The multipole moments for a ring with radius $r(\theta)$ are
$c_0 = (2\pi)^{-1} \int_0^{2\pi} r(\theta)d\theta$ for the monopole and
$c_m = \pi^{-1} \int_0^{2\pi} r(\theta) \cos m\theta d\theta$ and
$s_m = \pi^{-1} \int_0^{2\pi} r(\theta) \sin m\theta d\theta$ for $m>0$. }
then the uncertainty in the individual components is $\sigma^2=2\sigma_r^2/N$.  The number of
independent measurements is set by the resolution of the observations.  For a PSF with
full-width at half-maximum $FWHM$, the number of independent measurements is approximately
$N\simeq 2\pi b/ FWHM \simeq 50$ for HST observations of a lens with a critical radius
of $b=1\farcs0$.  The radius of the ring can be determined to accuracy
$\sigma_r \simeq FWHM/\sqrt{SNR}$ where $SNR$ is the signal-to-noise ratio of the ring
averaged over the resolution element.  Hence the multipole moments can be determined to
accuracy $\sigma/b \simeq 0.005 (10 FWHM/b)^{3/2} (10/SNR)^{1/2}$ for typical
HST observations.
If we compare this to the terms appearing in the shape of the Einstein ring, 
the accuracy with which the ellipticity and shear of the
lens can be determined from the shape of the ring is $\sigma(e_l)\simeq 0.02$ and
$\sigma(\gamma)\simeq 0.005 $ for the nominal noise level.  The axis ratio of the source
can be determined to an accuracy $\sigma(e_s)\sim 0.05$ from the higher order terms.

\section{Examples}

Our model for Einstein rings works well on synthetic data generated using an ellipsoidal
source.  The key question, however, is whether it works on real Einstein rings.  Here 
we illustrate our results for the Einstein ring images of the host galaxies in the lenses
PG~1115+080 (Impey et al. 1998), B~1608+656 (Fassnacht et al. 1996) and B~1938+666
(King et al. 1997).  We use the CfA/Arizona Space Telescope Lens Survey (CASTLES)
H-band images of the lenses (see Falco et al. 2000), which are shown in Figure 2.  Figure
3 shows the brightnesses of the rings as a function of azimuth, which are used to 
determine the positions of the extrema.  The maxima correspond to the centers of the
host galaxies, and the minima correspond to the points where the critical line crosses
the Einstein ring.  Our purpose is to illustrate the Einstein ring fitting method, with 
detailed treatments and discussions of $H_0$ for the individual systems being deferred 
to later studies.

\begin{figure}
\centerline{\psfig{figure=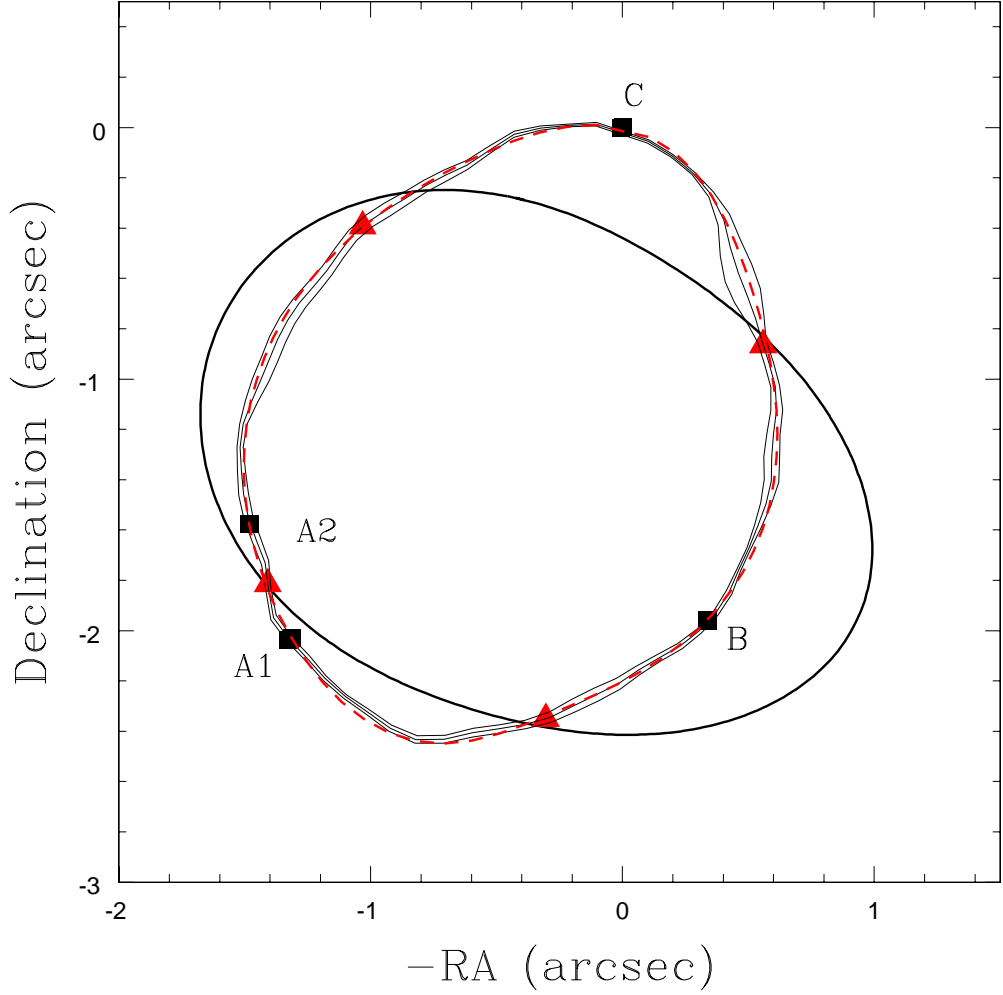,height=5.5in}}
\caption{PG~1115+080.  The quasar images A1, A2, B and C  are marked by the black points.
  The light black lines show the ring centroid and its uncertainties
  and the red/gray triangles mark the flux minima in the ring.  The red/gray dashed
  line is the best fit model
  of the ring and the heavy solid line is the tangential critical line of the best fit model.  
  The model was not constrained to fit the critical line crossings (see text).
  }
\end{figure}

PG~1115+080 (Weymann et al. 1980) was the second lens discovered. It consists of 4 images of a
$z_s=1.72$ quasar lensed by a $z_l=0.31$ early-type galaxy in small group (Tonry 1998, Kundic et al.
1997).  Detailed models of the system to interpret the time delay measurement (Schechter et al. 1997, 
also Barkana 1997) found a degeneracy in the models between the radial mass profile of the lens galaxy and
the value of the Hubble constant (see Keeton \& Kochanek 1997, Courbin et al. 1998,
Impey et al. 1998, William \& Saha 2000, Zhao \& Pronk 2000).  Impey et al. (1998) 
also discovered an Einstein ring formed from the host galaxy of the quasar, and could show that the
shape was plausibly reproduced by their models.  To extract the ring curve we first subtracted 
most of the flux from the quasar images to minimize the flux gradients along the ring.
Figure 4 schematically illustrates the positions of the four quasars, the location of the Einstein
ring, and its uncertainties.  We fit the data using an SIE for the primary lens galaxy and a singular
isothermal sphere for the group to which the lens belongs, which we already knew would provide a
statistically acceptable fit to all properties of the quasar images but the peculiar A1/A2 flux
ratio (see Impey et al. 1998).  We forced the images of the center of the host galaxy to be
within 10~mas of the quasar images, which tightly constrains the host position without introducing
a significant double-counting of the quasar position constraints.  We monitored, but did not include,
the critical line crossing constraints in the fits.
The Impey et al. (1998) model naturally reproduces the Einstein ring with a flattened host galaxy
($b/a=0.58\pm0.02$, PA$=-17^\circ\pm2^\circ$) centered on the quasar. 
All subsequent lens models had negligible 
changes in the shape and orientation of the source.  The fit to the ring is somewhat worse than
expected (a $\chi^2$ per ring point of $1.3$), in part due to residual systematic
problems from subtracting the point sources such as the ripple in the ring between the bright
A1 and A2 quasar images (see Figure 4).  Figure 4 also shows the locations of the 
flux minima in the ring, which should correspond to the points where the critical line crosses
the ring.  Three of the four flux minima lie where the model critical line crosses the ring given
the uncertainties.  The Northeast flux minimum, however, agreed with no attempted model. 

We examined whether the constraints from the Einstein ring can break the degeneracy between the
mass profile and the Hubble constant in two steps.  We first simulated the system with synthetic
data matching our best fit to the quasars and the Einstein ring using an SIE lens model.  We fit
the synthetic data using an ellipsoidal pseudo-Jaffe model for the lens instead of an SIE.  In 
previous models of the system we had found $H_0=45\pm4$~km~s$^{-1}$~Mpc$^{-1}$ for the SIE
model (the limit where the pseudo-Jaffe break radius $a\rightarrow\infty$, see \S2.2), and
$H_0=65\pm5$~Mpc$^{-1}$ for $a\simeq1\farcs0$ where the mass is as centrally concentrated
as the lens galaxy's light.  For the synthetic data the $\chi^2$ statistic shows a significant
rise for $a < 2\farcs0$ when the Einstein ring constraints are included, with roughly
equal contributions from the position of the ring, the positions of the flux maxima, and the
positions of the flux minima/critical line crossings.  In the second step we fit the real
data using the same models, but without the constraints on the locations of the critical
line crossings.  The rise in the $\chi^2$ statistic as we reduce the break radius $a$
is similar to that found for the synthetic data, and our 2$\sigma$ upper bound on
the break radius of the pseudo-Jaffe model is $a > 2\farcs0$.  The mass distribution of
the lens galaxy cannot be more centrally concentrated than the Einstein ring, and for the
Schechter et al. (1997) time delays it requires a low value of the Hubble constant,
$H_0 < 60$~km~s$^{-1}$~Mpc$^{-1}$.

\begin{figure}
\centerline{\psfig{figure=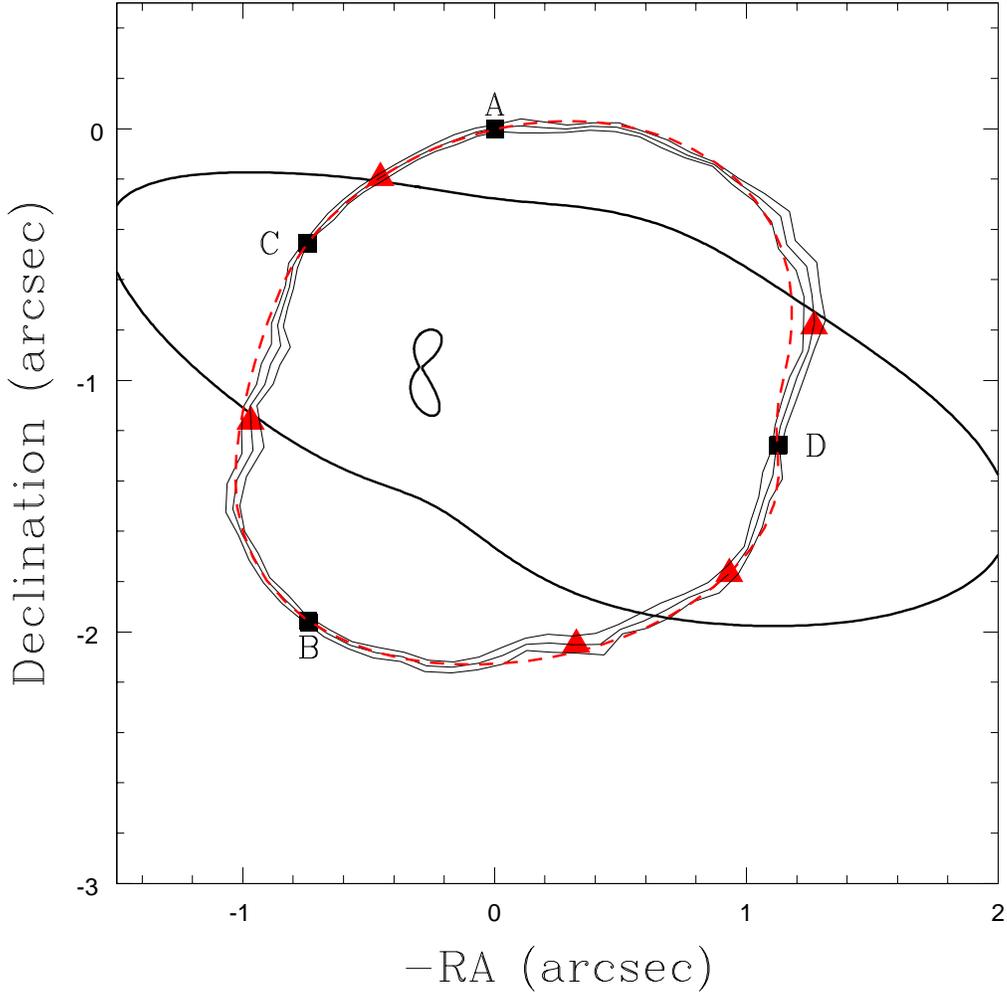,height=5.5in}}
\caption{B~1608+656.  The radio images A, B, C and D are aligned with the infrared point sources
  and are marked by the black points.
   The light black lines show the ring centroid and its uncertainties
  and the red/gray triangles mark the flux minima in the ring.  The red/gray dashed line is the best fit model
  of the ring and the heavy solid line is the tangential critical line of the best fit model.
  The model was not constrained to fit the critical line crossings (see text).   The two flux
  minima in the Southwestern ring quadrant are an artifact of the noise in the profile 
  (see Figure 3). }
\end{figure}
 
B~1608+656 (Myers et al. 1995, Fassnacht et al. 1996) consists of four unresolved radio images of a 
$z_s=1.394$ radio galaxy created by a lens consisting of two $z_l=0.63$ galaxies which lie inside
an optical and infrared Einstein ring image of the host galaxy of the AGN.  Fassnacht et al. 
(1999, 2000) have measured all three relative time delays for the radio images, which were
interpreted to determine a Hubble constant by Koopmans \& Fassnacht (1999).  We fit the 
system with two SIE models for the two galaxies enclosed by the ring, and an external shear
to represent the local environment or other perturbations.  Koopmans \& Fassnacht (1999)
added large core radii to their models to avoid creating 7 rather than 4 images.  Figure 5 
shows the ring and our fit.  In our models,
we avoid producing an unobserved image lying between the two galaxies by having a significantly
larger mass ratio between the two galaxies than in the Koopmans \& Fassnacht (1999) models. In
fact, the mass ratio we find in our fits is relatively close to the observed flux ratio of
the two galaxies.  The fit to the ring is not as clean as in PG~1115+080 although the goodness
of fit is reasonably good (an average $\chi^2$ per ring point of $1.3$).  The overall boxy
shape of the ring is reproduced by a source with an axis ratio of $0.69\pm0.02$ and a 
major axis PA of $-40^\circ\pm4^\circ$. The largest deviations of the model from the
data lie on an East-West axis where galaxy subtraction errors or dust in the lens galaxies
(Blandford et al. 2000) would most affect our extraction of the ring.  The locations of the
flux minima are broadly consistent with the model (see Figure 5), although the measurement
accuracy is poor.  In the Southwestern quadrant we found two minima in the ring flux, 
which is an artifact created by the noisy, flat brightness profile in that quadrant (see Figure 3). 

\begin{figure}
\centerline{\psfig{figure=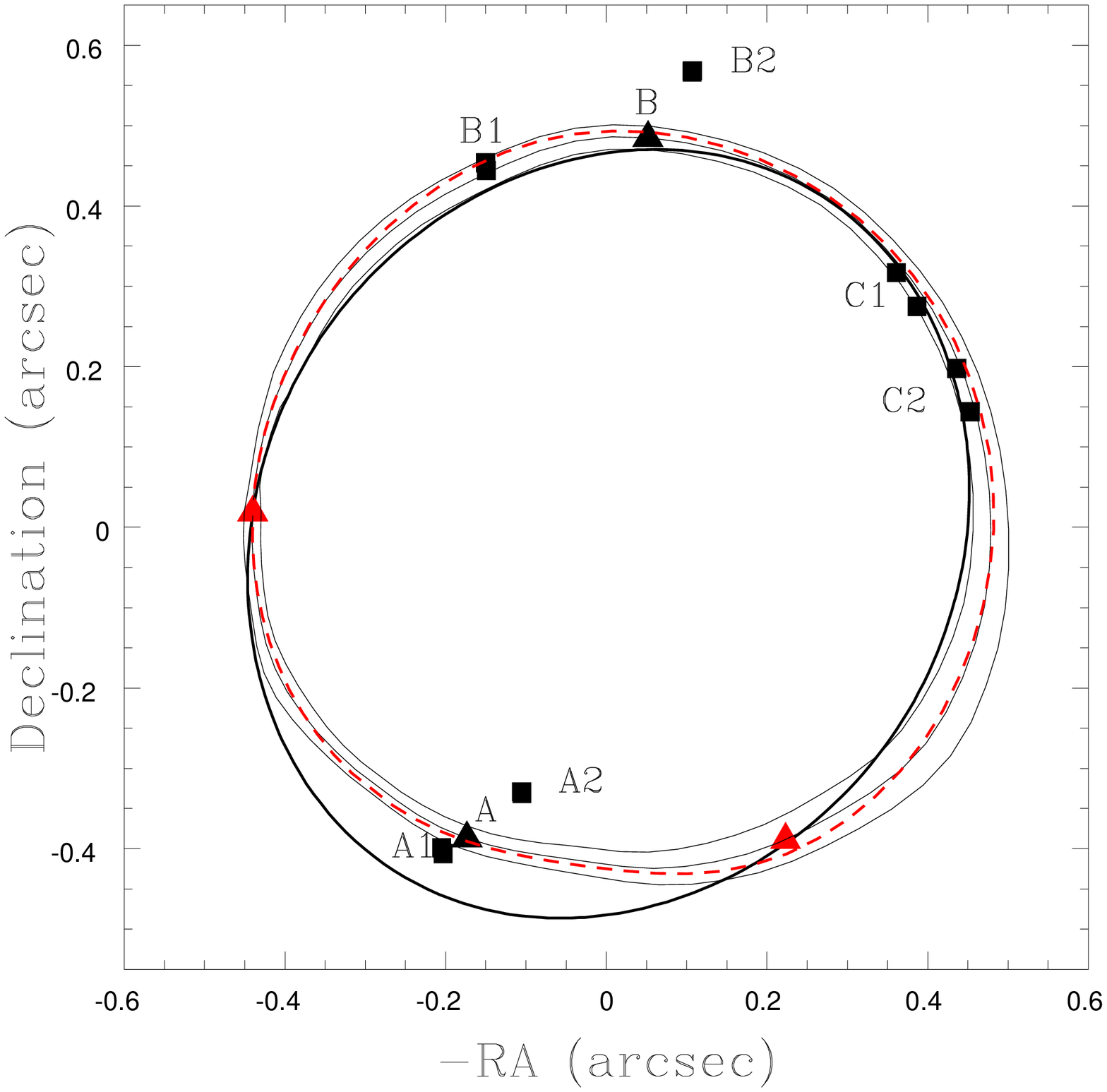,height=5.5in}}
\caption{B~1938+666.  The radio images A1, B1, C1 and C2 form a four-image system and the
 radio images A2 and B2 form a two-image system (black squares). There are two images of 
 the center of the host galaxy A and B (black triangles).  There are two points for the
 C1 and C2 images corresponding to the observed and model positions for the images.  
  The light black lines show the ring centroid and its uncertainties
  and the red/gray triangles mark the flux minima in the ring.  The red/gray dashed line 
  is the best fit model
  of the ring and the heavy solid line is the tangential critical line of the best fit model.
  The model was constrained to fit the critical line crossings (see text), and the alignment
  of the optical and radio data was optimized as part of the model.  
     }
\end{figure}

B~1938+666 (King et al. 1997) has a two-image system (A2/B2) and a four-image system (A1/B1/C1/C2)
of compact radio sources (see Figure 6).  The B1, C1 and C2 images are connected by a low surface 
brightness radio arc.  The lens galaxy is a normal early-type galaxy at $z_l=0.88$ (Tonry \& Kochanek 2000), 
but the source redshift is unknown.  Surrounding the lens galaxy is an almost perfectly circular infrared
Einstein ring image of the host galaxy (King et al. 1998, see Figure 2).  The radio sources are
multiply-imaged lobes rather than a central core, so the center of the galaxy is not aligned with
the doubly imaged radio source.  We fit the 6 compact components and the infrared ring,
allowing the model to optimize the registration of the radio and infrared data. 
The lens is very symmetric and the source is very close to the lens center, so we cannot
measure the axis ratio of the source to high precision.  We find an axis ratio of $0.62\pm0.14$
and a major axis position angle of $11^\circ\pm11^\circ$.  Since the ring data are very clean
compared to PG~1115+080 and B~1608+656 in the sense that there are no bright point sources
or complicated lenses which must be subtracted before extracting the ring, we included the
flux minima as a constraint on the models.  The fit to the ring is too good given the 
formal errors (for the overall trace we find a $\chi^2$ per point of $0.2$, a $\chi^2$ per 
flux peak of $0.3$, and a $\chi^2$ per flux minimum of $0.2$).  From our experience with
PG~1115+080 and B~1608+656 we had settled on a very conservative error model for the ring
properties, which is too pessimistic for B~1938+666 where it is so much easier to measure the
ring properties.  The VLBI component positions were fit less well (a $\chi^2$ per component
coordinate of $4.0$), in part because the accuracy of the component identifications for the 
arc images C1 and C2 are probably worse than their formal uncertainties.  The fit is
illustrated in Figure 6.  

\section{Summary}

Modern observations of gravitational lenses, both in the radio and with HST, routinely 
find extended lensed structures.  In particular, Einstein ring images of the host galaxies 
of the lensed sources (AGN, radio-loud and radio-quiet quasars) are frequently detected 
even in short infrared images of gravitational lenses.  The discovery
of these additional images is critical to expanding the use of lenses to determine the
mass distributions of galaxies and the Hubble constant, where the models have been
limited by the finite number of constraints supplied by the two or four images of the 
active nucleus.  

We developed a simple theoretical model for Einstein rings and demonstrated it using 
the rings observed in PG~1115+080, B~1608+656 and B~1938+666.  The assumption of
ellipsoidal symmetry for the source works well, and we can accurately and simultaneously
determine the shapes of the host galaxy and the lens potential.  Contrary to popular
belief, the distortions introduced by lensing are not a major complication to studying 
the properties of the host galaxy.  In fact,  host galaxies and gravitational lensing 
provide a virtuous circle.  The magnification by the
lens enormously improves the contrast between the host galaxy and the central engine 
over unlensed host galaxies (by factors of $10$--$10^3$ in surface brightness contrast!).
This makes it significantly easier to find and analyze the host galaxies of high redshift quasars.
In return, the host galaxy provides significant additional constraints on the mass distribution 
of the lens.  These constraints break the lens model degeneracies which have made it difficult
to determine the mass distribution of the lens or the Hubble constant from time delay
measurements.  Since we expect all quasars and AGN to have host galaxies, we can obtain these
additional constraints for any system where they are needed.


Our tests of the method worked extraordinarily well even though the existing observations of the lensed 
host galaxies were incidental to observations designed to efficiently find and study the lens galaxy
(see Falco et al. 2000).  The observations were short and could not afford the high overheads for
obtaining contemporaneous empirical PSFs.  There are now six gravitational lenses with time delay
measurements (B~0218+357, Biggs et al. 1999; Q~0957+561, Schild \& Thomson 1995, Kundic et al. 1997,
Haarsma et al. 1999; PG~1115+080, Schechter et al. 1997, Barkana 1997; B~1600+434, Koopmans et al. 2000, 
Hjorth et al. 2000; B~1608+656, Fassnacht et al. 1999, 2000; and PKS~1830--211, Lovell et al. 1998)
and lensed images of the host galaxy have been found in four of the six (Q~0957+561, Bernstein et al. 1997,
Keeton et al. 2000; PG~1115+080, Impey et al. 1998; B~1600+434, Kochanek et al. 1999; and 
B~1608+656, Koopmans \& Fassnacht 1999).  A dedicated program of longer observations with empirical
PSF data would be revolutionary and lead to a direct, accurate determination of the global value of 
the Hubble constant.

\bigskip
\noindent Acknowledgements:  The authors thank E. Falco for his comments.  Support for the CASTLES 
project was provided by NASA through grant numbers GO-7495 and GO-7887 from the Space Telescope
Science Institute, which is operated by the Association of Universities for Research in Astronomy,
Inc.  CSK and BAM are supported by the Smithsonian Institution. CSK is also supported by NASA
Astrophysics Theory Program grants NAG5-4062 and NAG5-9265.
\bigskip

\appendix
\section{An Analytic Model}

\def\ve{\hat{e}}
\def\vt{\vec{t}}
\def\vh{\vec{h}}
\def\tg{{\bf G}}

In this section we derive an analytic solution of eqn. (7) for the Einstein ring radius as a function
of azimuth for generalized isothermal potentials, $\phi=r r_0 F(\theta)$, in an external shear
field, where $r_0$ is a constant radial scale and we define $\langle F(\theta)\rangle = 1$
(see Zhao \& Pronk 2000, Witt, Mao \& Keeton 2000).  The model includes the SIE as a subcase.  
After defining the radial and tangential 
unit vectors, $\ve_r=d\vx/d\lambda$ and $\ve_\theta$, we find that the source plane tangent 
vector is
\begin{equation}
   \vt = \mt^{-1} \cdot \ve_r = \tg \cdot \ve_r 
\end{equation}
where 
\begin{equation}
   \tg =  \left(
             \begin{array}{cc}
                  1-\gamma_c  &-\gamma_s  \\
                  -\gamma_s   &1+\gamma_c  
             \end{array}
             \right)
\end{equation}
defines the external shear.  If there is no external shear, $\tg$ becomes the identity matrix and 
$\vt=\ve_r$.  The distance of the curve from the center of the ellipsoidal source is 
\begin{equation}
      \vu-\vu_0 = r \vt - r_0 F(\theta) \ve_r - r_0 F'(\theta) \ve_\theta - \vu_0
\end{equation}
where $F'(\theta)=dF/d\theta$. The radius of the Einstein ring relative to the lens is simply
\begin{equation}
       r =  { r_0\vh \cdot \st \cdot \vt + \vu_0 \cdot \st \cdot \vt \over
                   \vt \cdot \st \cdot \vt }.
\end{equation}
where $\vh=F \ve_r + F' \ve_\theta$.
Note that there is no transformation of the source shape which will eliminate any other variable 
(the lens potential, the external shear or the source position) from the shape of the ring, which
means that there are no simple parameter degeneracies between the potential and the source.
If there is no external shear ($\vt=\ve_r$), then the solution simplifies to
\begin{equation}
       r = r_0 F(\theta) + r_0 F'(\theta) { \ve_\theta \cdot \st \cdot \ve_r \over \ve_r \cdot \st \cdot \ve_r  }
              + { \vu_0 \cdot \st \cdot \ve_r \over \ve_r \cdot \st \cdot \ve_r }
         \rightarrow r_0 F(\theta) + \vu_0 \cdot \ve_r
\end{equation}
if the source is circular and $\st={\bf I}$. For comparison, the tangential critical radius of the
lens is 
\begin{equation}
       r_{crit } = r_0 \left[ F(\theta)+F''(\theta)\right]  \left[ { \ve_r \cdot \tg \cdot \ve_r \over |\tg|} \right]
\end{equation}
We have found no general analytic solutions for models with a different radial dependence for the density, but
limited analytic progress can be made for potentials of the form $\phi = r^\beta r_0^{2-\beta} F(\theta)$
with the source at the origin.  The equation for the ring location is then a quadratic in 
$W=(r/r0)^{2-\beta}$,
\begin{equation}
     0 = W^2 \vt \cdot \st \cdot \vt - \beta W \vh \cdot \st \cdot \vt + (\beta-1)\vh\cdot\st\cdot\vh
\end{equation}
where the definition of $\vh$ is modified to be $\vh = \beta F \ve_r + F' \ve_\theta$.  
The result is of limited use because 
the source offsets are important in explaining the observed ring shapes.  The radial profile
clearly modifies the ring shape, but numerical experiments are required to determine the ability of the
ring shape to discriminate between radial mass profiles given the freedom to adjust the source shape and
position, the lens shape and the external shear.

\end{document}